\documentclass[twocolumn]{aastex63}
\usepackage{lipsum}
\usepackage{mathtools}
\usepackage{amsmath}
\usepackage{subfigure}
\usepackage{amssymb}

\received{April 22, 2020}
\revised{June 4, 2020}
\accepted{June 9, 2020}

\shorttitle{Cosmic MRD from young star clusters}
\shortauthors{Santoliquido et al.}
\graphicspath{{./}{figures/}}

\begin{document}

\title{The cosmic merger rate density evolution of compact binaries formed in young star clusters and in isolated binaries}

\correspondingauthor{Filippo Santoliquido}
\email{filippo.santoliquido@phd.unipd.it}

\author[0000-0003-3752-1400]{Filippo Santoliquido}
\affiliation{Physics and Astronomy Department Galileo Galilei, University of Padova, Vicolo dell'Osservatorio 3, I--35122, Padova, Italy}
\affiliation{INFN--Padova, Via Marzolo 8, I--35131 Padova, Italy}

\author[0000-0001-8799-2548]{Michela Mapelli}
\affiliation{Physics and Astronomy Department Galileo Galilei, University of Padova, Vicolo dell'Osservatorio 3, I--35122, Padova, Italy}
\affiliation{INFN--Padova, Via Marzolo 8, I--35131 Padova, Italy}
\affiliation{INAF--Osservatorio Astronomico di Padova, Vicolo dell'Osservatorio 5, I--35122, Padova, Italy}

\author[0000-0003-3462-0366]{Yann Bouffanais}
\affiliation{Physics and Astronomy Department Galileo Galilei, University of Padova, Vicolo dell'Osservatorio 3, I--35122, Padova, Italy}
\affiliation{INFN--Padova, Via Marzolo 8, I--35131 Padova, Italy}

\author[0000-0002-8339-0889]{Nicola Giacobbo}
\affiliation{Physics and Astronomy Department Galileo Galilei, University of Padova, Vicolo dell'Osservatorio 3, I--35122, Padova, Italy}
\affiliation{INFN--Padova, Via Marzolo 8, I--35131 Padova, Italy}

\author[0000-0003-2654-5239]{Ugo N. Di Carlo}
\affiliation{INFN--Padova, Via Marzolo 8, I--35131 Padova, Italy}
\affiliation{Dipartimento di Scienza e Alta Tecnologia, University of Insubria, Via Valleggio 11, I-22100 Como, Italy}

\author[0000-0002-5699-5516]{Sara Rastello}
\affiliation{Physics and Astronomy Department Galileo Galilei, University of Padova, Vicolo dell'Osservatorio 3, I--35122, Padova, Italy}
\affiliation{INFN--Padova, Via Marzolo 8, I--35131 Padova, Italy}

\author[0000-0003-0570-785X]{M. Celeste Artale}
\affiliation{Institut f\"{u}r Astro- und Teilchenphysik, Universit\"{a}t Innsbruck, Technikerstrasse 25/8, 6020 Innsbruck, Austria}

\author[0000-0003-4893-2993]{Alessandro Ballone}
\affiliation{Physics and Astronomy Department Galileo Galilei, University of Padova, Vicolo dell'Osservatorio 3, I--35122, Padova, Italy}
\affiliation{INFN--Padova, Via Marzolo 8, I--35131 Padova, Italy}

\begin{abstract}
Next generation ground-based gravitational-wave detectors will observe binary black hole (BBH) mergers up to redshift $z\gtrsim{}10$, probing the evolution of compact binary (CB) mergers across cosmic time. Here, we present a new data-driven model to estimate the cosmic merger rate density (MRD) evolution of CBs, by coupling catalogs of CB mergers  with observational constraints on the cosmic star formation rate density and on the metallicity evolution of the Universe. We adopt catalogs of CB mergers derived from recent $N-$body and population-synthesis simulations, to describe the MRD of CBs formed in young star clusters (hereafter, dynamical CBs) and in the field (hereafter, isolated CBs). The local MRD of dynamical BBHs is $\mathcal{R}_{\rm BBH}=64^{+34}_{-20}$ Gpc$^{-3}$   yr$^{-1}$,  consistent with the 90\% credible interval from the first and second observing run (O1 and O2) of the LIGO--Virgo collaboration, and with the local MRD of isolated BBHs ($\mathcal{R}_{\rm BBH}=50^{+71}_{-37}$ Gpc$^{-3}$  yr$^{-1}$). 
The local MRD of dynamical and isolated  black hole -- neutron star binaries is  $\mathcal{R}_{\rm BHNS}=41^{+33}_{-23}$ and $49^{+48}_{-34}$~Gpc$^{-3}$  yr$^{-1}$, respectively. Both values are consistent with the upper limit inferred from O1 and O2. Finally, the local MRD of dynamical binary neutron stars (BNSs, $\mathcal{R}_{\rm BNS}=151^{+59}_{-38}$ Gpc$^{-3}$  yr$^{-1}$) is a factor of two lower than the local MRD of isolated BNSs ($\mathcal{R}_{\rm BNS}=283^{+97}_{-75}$ Gpc$^{-3}$  yr$^{-1}$).
The MRD for all CB classes grows with redshift, reaching its maximum at $z \in [1.5,2.5]$, and then decreases. This trend springs from the interplay between cosmic star formation rate, metallicity evolution and delay time of binary compact objects. 
\end{abstract}


\section{Introduction} \label{sec:intro}
Thirteen gravitational-wave (GW) events have been published by the LIGO--Virgo collaboration (LVC, \citealt{LIGOdetector,VIRGOdetector}) since 2016, eleven of them associated with binary black hole (BBH) mergers \citep{abbottGW150914,abbottGW151226,abbottO1,abbottGW170814,abbottGW170104,abbottGW170608,abbottO2,abbottO2popandrate,abbottGW190412} and two events with binary neutron stars (BNSs) \citep{abbottGW170817,abbottGW190425}. Several additional  BBHs were claimed by other studies, based on different pipelines \citep{2019PhRvD.100b3011V,2019arXiv190407214V,2019arXiv191009528Z,2019PhRvD.100b3007Z}. This data sample marks the dawn of GW astrophysics, and makes it possible to estimate the local merger rate density (MRD) of binary compact objects. The LVC has inferred a local MRD (within 90~$\%$ credible intervals) $\mathcal{R}_{\rm BBH}\sim{}24-140$~Gpc$^{-3}$~yr$^{-1}$ \citep{abbottO2popandrate}, $\mathcal{R}_{\rm BHNS}<610$~Gpc$^{-3}$~yr$^{-1}$ \citep{abbottO2}
and  $\mathcal{R}_{\rm BNS}=250-2810$~Gpc$^{-3}$~yr$^{-1}$ \citep{abbottGW190425} for BBHs, black hole--neutron star binaries (BHNSs) and BNSs, respectively. 

At design sensitivity, LIGO and Virgo will be sensitive to BBHs up to $z\gtrsim{}1$ and to BNSs up to $z\sim{}0.1$. Moreover, third-generation ground-based GW interferometers, Einstein Telescope in Europe \citep{punturo2010,maggiore2020} and Cosmic Explorer in the US \citep{reitze2019}, are being planned, with a target sensitivity that will allow us to observe BBH mergers up to $z\gtrsim{}10$  and BNS mergers up to $z\sim{}2$ \citep{kalogera2019}. This will open new perspectives on the study of binary compact objects: we might even reconstruct their formation channels through their redshift evolution. Moreover, we will be able to infer their delay time (i.e. the time elapsed from their formation to their merger, \citealt{2019ApJ...878L..12S,2019ApJ...878L..13S}) and we might constrain the cosmic star formation rate (SFR) and metallicity evolution based on GWs \citep{kalogera2019}. 
Hence, it is crucial to model the cosmic evolution of binary compact objects.

Current theoretical predictions about the cosmic MRD follow two approaches. The first one  consists in seeding  compact-object binaries (CBs) in cosmological simulations, based on the properties of simulated galaxies \citep{lamberts2016,lamberts2018,oshaughnessy2017,schneider2017,mapelli2017,mapelli2018,mapelli2018b,mapelli2019,toffano2019,artale2019a,artale2020b,artale2020}. This approach is effective if we are interested in the properties of the host galaxies, but is computationally challenging. The alternative approach consists in interfacing catalogs from population-synthesis models, or simpler phenomenological models, with data-driven prescriptions for the evolution of the star-formation rate and the metallicity in the Universe \citep{oshaughnessy2010,dominik2013,dominik2015,belczynski2016,giacobbo2018b,giacobbo2020,baibhav2019,neijssel2019,boco2019,tang2020}. The latter approach  is more effective to sample the parameter space and can be used to probe different formation pathways (such as the isolated binary formation and the dynamical formation scenarios).

While the aforementioned studies focus only on the formation of CBs from isolated binary evolution, several additional works have tried to quantify the MRD evolution of BBHs from globular clusters \citep{portegieszwart2000,tanikawa2013,rodriguez2016,askar2017,fragionekocsis2018,choksi2018,choksi2019,hong2018,rodriguezloeb2018}, nuclear star clusters \citep{antonini2016,petrovich2017,2020ApJ...891...47S}, AGN disks \citep{2017ApJ...835..165B,2017MNRAS.464..946S,mckernan2018,2019ApJ...876..122Y,2019arXiv191208218T}  and open clusters \citep{ziosi2014,kumamoto2020}. Among these studies, \cite{rodriguezloeb2018} compared the MRD estimated for isolated binaries with the one inferred for globular clusters. 

No previous study focused on the cosmic MRD of BBHs born in young star clusters. Since the majority of massive stars are thought to be born in young star clusters, these are a crucial environment for binary compact objects, at least in the local Universe \citep{lada2003,portegieszwart2010}.  Young star clusters are short-lived (few Myr to few Gyr) and generally less massive than globular clusters, but are much more common. They continuously form across cosmic time (both at high and at low redshift), while globular cluster formation is strongly suppressed at low redshift. As in globular clusters, dynamical encounters affect the formation of CBs in young star clusters, but with two crucial differences: i) the two-body relaxation timescale is at least a factor of ten shorter in young star clusters with respect to globular clusters, ii) the escape velocity from a typical young star cluster is a factor of $5-10$ lower than that from a globular cluster \citep{portegieszwart2010}. Hence, most dynamical encounters in young star clusters happen in the first $\sim{}10$ Myr and involve the stellar progenitors of a binary compact object, rather than the binary compact object itself \citep{mapelli2016,kumamoto2019,dicarlo2019a,dicarlo2019b}. After this early dynamical interaction phase, binary compact objects are generally ejected from their parent young star cluster.

Here, we derive the MRD of CBs (BBHs, BHNSs and BNSs) from young star clusters and compare it with the prediction from isolated binary evolution, using a new data-driven approach. We combine catalogs of simulated CB mergers with the cosmic SFR density evolution inferred by \cite{madau2017} and with a description of the metallicity evolution based on measurements of damped Lyman$-\alpha$ systems up to redshift $z\sim{}5$ \citep{decia2018}. The catalogs of simulated mergers of CBs formed in young star clusters (hereafter, dynamical CBs) come from the $N-$body simulations presented in \cite{rastello2020} and \cite{dicarlo2020}, while the isolated CBs are taken from \cite{giacobbo2018b}.




\section{Methods} \label{sec:methods}

\subsection{Cosmic MRD}
\label{sec:mrd}

We derive  the cosmic MRD of CBs as 
\begin{eqnarray}
\label{eq:mrd}
   \mathcal{R}(z) = \frac{\rm d\quad{}\quad{}}{{\rm d}t_{\rm lb}(z)}\int_{z_{\rm max}}^{z}\psi(z')\,{}\frac{{\rm d}t_{lb}(z')}{{\rm d}z'}\,{}{\rm d}z'\times{}\nonumber{}
   \\
   \int_{Z_{\rm min}(z')}^{Z_{\rm max}(z')}\eta{}(Z)\,{}\mathcal{F}(z',z, Z)\,{}{\rm d}Z
\end{eqnarray}
where $t_{\rm lb}(z)$ is the look-back time at redshift $z$, $\psi(z')$ is the cosmic SFR density at redshift $z'$, $Z_{\rm min}(z')$ and $Z_{\rm max}(z')$ are the minimum and maximum metallicity of stars formed at redshift $z'$, $\eta{}(Z)$ is the merger efficiency at metallicity $Z$, and $\mathcal{F}(z', z, Z)$ is the fraction of CBs that form at redshift $z'$ from stars with metallicity $Z$ and merge at redshift $z$, normalized to all CBs that form from stars with metallicity $Z$. To calculate the lookback time we take the cosmological parameters ($H_{0}$, $\Omega_{\rm M}$ and $\Omega_{\Lambda}$)  from \cite{planck2016}. The maximum considered redshift in equation~\ref{eq:mrd} is $z_{\rm max}=15$, which we assume to be the epoch of formation of the first stars.

The cosmic SFR density $\psi(z)$ is given by the following fitting formula \citep{madau2017}
\begin{equation}
\label{eq:sfrd}
 \psi(z) = 0.01\,{}\frac{(1+z)^{2.6}}{1+[(1+z)/3.2]^{6.2}}~\text{M}_\odot\,{}\text{Mpc}^{-3}\,{}\text{yr}^{-1}.
\end{equation}
To estimate the uncertainty on $\psi{}(0)$, we assume that the errors follow a log-normal distribution with mean $\log{\psi{}(0)}=-2$ and standard deviation $\sigma_{\log{\psi}}=0.2$  (taking into account the typical $1\,{}\sigma{}$ error bars on single data points, see Figure~9 of \citealt{madaudickinson2014}).

 We define the merger efficiency $\eta(Z)$ as
\begin{equation}
    \eta (Z) = \frac{\mathcal{N}_{\text{TOT}}(Z)}{M_\ast{}(Z)},
\end{equation}
where $\mathcal{N}_{{\rm TOT}}(Z)$ is the total number of CBs (BBHs, BHNSs or BNSs) that have delay time (i.e. the time elapsed from the formation of the binary star to the merger of the two compact objects) $t_{\rm del}\leq{}14$ Gyr  
born from stars with metallicity $Z$ in our population-synthesis simulations, and $M_\ast{}(Z)$ is the total initial stellar mass (corresponding to the zero-age main sequence mass) simulated with metallicity $Z$. Thus, the merger efficiency is the number of mergers occurring in a population of initial stellar mass $M_\ast{}$ and metallicity $Z$, integrated over a Hubble time (see e.g. \citealt{giacobbo2018b,klencki2018}). 

In equation~\ref{eq:mrd}, the values of $\eta{}(Z)$ and $\mathcal{F}(z', z, Z)$ are estimated from catalogs of CB mergers obtained with population synthesis and with dynamical simulations, as detailed in the next sections. The catalogs contain information on the masses of the two compact objects, the delay time  and the metallicity of the progenitor stars. In practice, since we have 6 (3) catalogs corresponding to 6 (3) different metallicities for isolated (dynamical) binary compact objects, the values of $\eta{}(Z)$ are linearly interpolated between the available metallicities (Figure~\ref{fig:eta}).

The value of $\mathcal{F}(z', z, Z)$ depends on the metallicity $Z$ of stars that form at redshift $z'$. To derive the average metallicity evolution as a function of redshift we use the following fitting formula:
\begin{equation}
\label{eq:decia}
 \mu(z) =  \log{ \left(\frac{Z(z)}{Z_\odot}\right)} = \log{a}\,{}+\,{}b\,{}z,
\end{equation}
where $a=1.04\pm{}\,{0.14}$ and $b=-0.24\pm{}0.14$. In the above equation, the slope $b$ 
comes from \cite{decia2018}, who provide a fit to the metallicity evolution of a large sample of damped Lyman$-\alpha$ systems with redshift between 0 and 5. 
The original fit by \cite{decia2018} yields a metallicity $Z(z=0)=0.66$ Z$_\odot$, which is low compared to the average stellar  metallicity measured at redshift zero (see, e.g., the discussion in \citealt{madaudickinson2014}). Hence, in equation~\ref{eq:decia}, we have re-scaled the fitting formula provided by \cite{decia2018} to yield $Z(z = 0) = (1.04\pm{}\,{0.14})~{\rm Z}_\odot$,  where $Z_\odot=0.019$, consistent with the average metallicity of  galaxies at $z\sim{}0$ from the Sloan Digital Sky Survey \citep{gallazzi2008}. The value of $a=1.04\pm{}0.14$ adopted in equation~\ref{eq:decia} is the result of this rescaling. The quoted uncertainties on both $a$ and $b$ are at 1 $\sigma{}$, assuming (as done in the original papers by \citealt{gallazzi2008} and \citealt{decia2018}) that the observational values follow a Gaussian distribution.

We model the distribution of stellar metallicities $\log{(Z/{\rm Z}_\odot)}$ at a given redshift as a normal distribution with mean value $\mu{}(z)$ from eq.~\ref{eq:decia} and standard deviation\footnote{We assume $\sigma{}_{Z}=0.20$, based on the metallicity spread found in cosmological simulations (e.g., {\sc eagle}, \citealt{artale2019a}). In a companion paper, we discuss the impact of a different choice of $\sigma_Z$ (Santoliquido et al., in preparation; see also \citealt{chruslinska2019,chruslinska2020}).}  $\sigma_{Z} = 0.20$ 
\begin{equation}
\label{eq:prob}
p(z', Z) = \frac{1}{\sqrt{2 \pi\,{}\sigma_{Z}^2}}\,{} \exp\left\{{-\,{} \frac{\left[\log{(Z/{\rm Z}_\odot)} - \mu(z')\right]^2}{2\,{}\sigma_{Z}^2}}\right\}.
\end{equation}{}

Based on our definition, $\mathcal{F}(z',z,Z)$ and  $p(z', Z)$ are connected by the following relation:
\begin{equation}
\mathcal{F}(z',z,Z)=\frac{\mathcal{N}(z,Z)}{\mathcal{N}_{\rm TOT}(Z)}\,{}p(z', Z),
\end{equation}
where $\mathcal{N}(z,Z)$ is the number of CBs that form from stars with metallicity $Z$ and merge at redshift $z$, while $\mathcal{N}_{\rm TOT}(Z)$ is the total number of CBs that merge within a Hubble time and form from  stars  with  metallicity $Z$ (as already detailed above).

We performed $10^{3}$ realizations of equation~\ref{eq:mrd} per each considered model, 
in order to estimate the impact of observational uncertainties on the MRD. At each realization, we randomly draw  the normalization value of the SFR density (equation~\ref{eq:sfrd}), the intercept and the slope of the average metallicity (equation~\ref{eq:decia}) from three  Gaussian distributions with mean (standard deviation) equal to $\log{\psi{}(0)}=-2$ ($\sigma_{\log{\psi}}=0.2$), $a=1.04$ ($\sigma_a=0.14$) and $b=-0.24$ ($\sigma_b=0.14$), respectively. The value of the intercept and that of the slope are drawn separately, assuming no correlation.
This procedure is implemented in the new python script {\sc cosmo}$\mathcal{R}${\sc ate}, which allows us to calculate up to $10^{3}$ models per day on a single core. -

\subsection{Population synthesis}

The catalogs of isolated binaries have been generated with our population-synthesis code {\sc mobse}  \citep{mapelli2017, giacobbo2018, giacobbo2018b, mapelli2018}.  In {\sc mobse}, the mass loss of massive hot stars  is described as $\dot M \propto Z^{\beta}$, where $\beta$ is defined as in \cite{giacobbo2018}:
  \begin{equation}\label{eq:gamma}
    \beta=
    \begin{cases}
      0.85, & \text{if}\ \Gamma_{e} \leq{} 2/3 \\
      2.45-2.4\Gamma_{e}, & \text{if}\  2/3<\Gamma_{e} \leq{} 1\\
      0.05, & \text{if}\ \Gamma_{e} > 1
    \end{cases}
  \end{equation}
In eq.~\ref{eq:gamma},   $\Gamma_e$ is the Eddington ratio, i.e. the ratio between the luminosity of the star and its Eddington value. 

{\sc mobse} includes two different prescriptions for core-collapse supernovae (SNe) from \cite{fryer2012}: the \textit{rapid} and the \textit{delayed} SN models. The former model assumes that the SN explosion is launched $\lesssim$ 250 ms after the bounce, while the latter has a longer timescale ($\gtrsim$ 500 ms). In both models, a star is assumed to directly collapse into a black hole (BH) if its final carbon-oxygen mass is $\gtrsim 11 ~M_\odot$. For the simulations described in this paper we adopt the rapid model, which enforces a gap in the mass function of compact objects between 2 and 5 M$_\odot$. Recipes for electron-capture SNe are included in {\sc mobse} as described in \cite{giacobbo2019}.

Prescriptions for  pair instability and pulsational pair instability are implemented using the fitting formulas derived by \cite{spera2017}. In particular, stars which grow a helium core mass $64 \leq m_{He}/{\rm M}_\odot \leq 135$ are completely disrupted by pair instability and leave no compact objects, while stars with $32 \leq m_{He}/{\rm M}_\odot < 64$ undergo a set of pulsations, which enhance mass loss and cause the final compact object mass to be significantly smaller than it would be if we had accounted only for core-collapse SNe.

Natal kicks are randomly drawn from a Maxwellian velocity distribution. In the run presented here, 
we adopt a one-dimensional root mean square velocity $\sigma = 15 ~\text{km}$~s$^{-1}$ for neutron stars. BH natal kicks are drawn from the same distribution as neutron-star kicks, but reduced by the amount of fallback as $v_{\text{KICK}} = (1 - f_{\rm fb})\,{}v$, where $f_{\rm fb}$ is the fallback parameter described in \cite{fryer2012} and $v$ is the velocity drawn from the Maxwellian distribution.

Binary evolution processes such as tidal evolution, Roche lobe overflow, common envelope and GW energy loss are taken into account as described in \cite{hurley2002}. In particular, the treatment of common
envelope is described by the efficiency parameter $\alpha$. In this work, we assume $\alpha = 5$, as suggested by recent studies \citep{fragos2019,giacobbo2020}. Orbital decay and circularization by GW emission are calculated according to \cite{peters1964}.

We have simulated $6\times{}10^7$ isolated binaries with {\sc mobse}, $10^7$ per each metallicity we considered ($Z=0.0002$, 0.0008, 0.002, 0.008, 0.016 and 0.02). The mass of the primary star is randomly drawn from a \cite{kroupa2001} initial mass function, with minimum mass 5 M$_\odot$ and maximum mass 150 M$_\odot$. The orbital periods, eccentricities and mass ratios of binaries are drawn from \cite{sana2012}. In particular, we derive the mass ratio $q=m_2/m_1$ as $\mathcal{D}(q) \propto q^{-0.1}$ with $q\in [0.1-1]$, the orbital period $P$ from $\mathcal{D}(\Pi) \propto \Pi^{-0.55}$ with $\Pi = \log_{10}(P/\text{day}) \in [0.15 -5.5]$ and the eccentricity $e$ from $\mathcal{D}(e) \propto e^{-0.42}~~\text{with}~~ 0\leq e \leq 1$. These simulations are part of run CC15$\alpha{}5$ in \cite{giacobbo2018b}. 

\subsection{Dynamics}
We derive the catalogs of CB mergers from a set of direct N-body simulations already described in \cite{dicarlo2020} and \cite{rastello2020}. These dynamical simulations were ran with the direct N-body code {\sc nbody6++gpu} \citep{wang2015,wang2016}, 
coupled with the population-synthesis code {\sc mobse}, as already described in \cite{dicarlo2019a}. In this way, the dynamical simulations include binary population synthesis, performed  with the same code as the isolated-binary simulations. 

The masses of the simulated young star clusters range from $ 300 ~{\rm M}_\odot$ to $30000~{\rm M}_\odot$. In particular, we consider $7.5\times{}10^4$ star clusters with mass $M_{\rm SC}\in{}[300,\,{}1000]$ M$_\odot$ ($2.5\times{}10^4$ runs per each considered metallicity: $Z=0.0002$, 0.002 and 0.02, from \citealt{rastello2020}) and 3000 star clusters with mass $M_{\rm SC}\in [1000,\,{}30000]$ M$_\odot$ (1000 runs per each considered metallicity: $Z=0.0002$, 0.002 and 0.02, presented as set~A in \citealt{dicarlo2020}). The total mass $M_{\rm SC}$ of a star cluster is drawn from a distribution $dN/dM_{\rm SC} \propto M_{\rm SC}^{-2}$, consistent with the mass function of young star clusters in the Milky Way \citep{lada2003}. 

 The initial half-mass radius $r_h$ of star clusters is distributed according to the Marks $\&$ Kroupa relation \citep{marks2012}, which relates the total mass of the  star cluster $M_{{\rm SC}}$ with its initial half mass radius $r_h$ as
\begin{equation}
  r_h = 0.10^{+0.07}_{-0.04} ~{\rm pc}\,{}\left(\frac{M_{{\rm SC}}}{{\rm M}_\odot}\right)^{0.13\pm 0.04}.
\end{equation}
  The star clusters are initialized in virial equilibrium.

The initial distribution of stellar positions and velocities in the star clusters have been generated through the {\sc mcluster} code \citep{kuepper2011}, according to a fractal distribution with fractal dimension $D=1.6$ \citep{goodwin2004}. This ensures that the initial conditions of the simulated star clusters are clumpy and asymmetric as observed embedded star clusters.  The mass of the stars is drawn from a \cite{kroupa2001} initial mass function between 0.1 and 150 M$_\odot$. The total initial binary fraction is  $f_{\rm bin}=0.4$. The mass ratios between secondary and primary star and the orbital properties of the binary systems (period and eccentricity) are drawn according to \cite{sana2012}, to ensure a fair comparison with the isolated binary simulations. The force integration includes a solar neighborhood-like static external tidal field. In particular, the simulated star clusters are assumed to be on a circular orbit around the center of the Milky Way with a semi-major axis of $8~\text{kpc}$ \citep{wang2016}. Each star cluster is evolved until its dissolution or for a maximum time $t = 100~\text{Myr}$.

\begin{figure}
\includegraphics[width = 0.45 \textwidth]{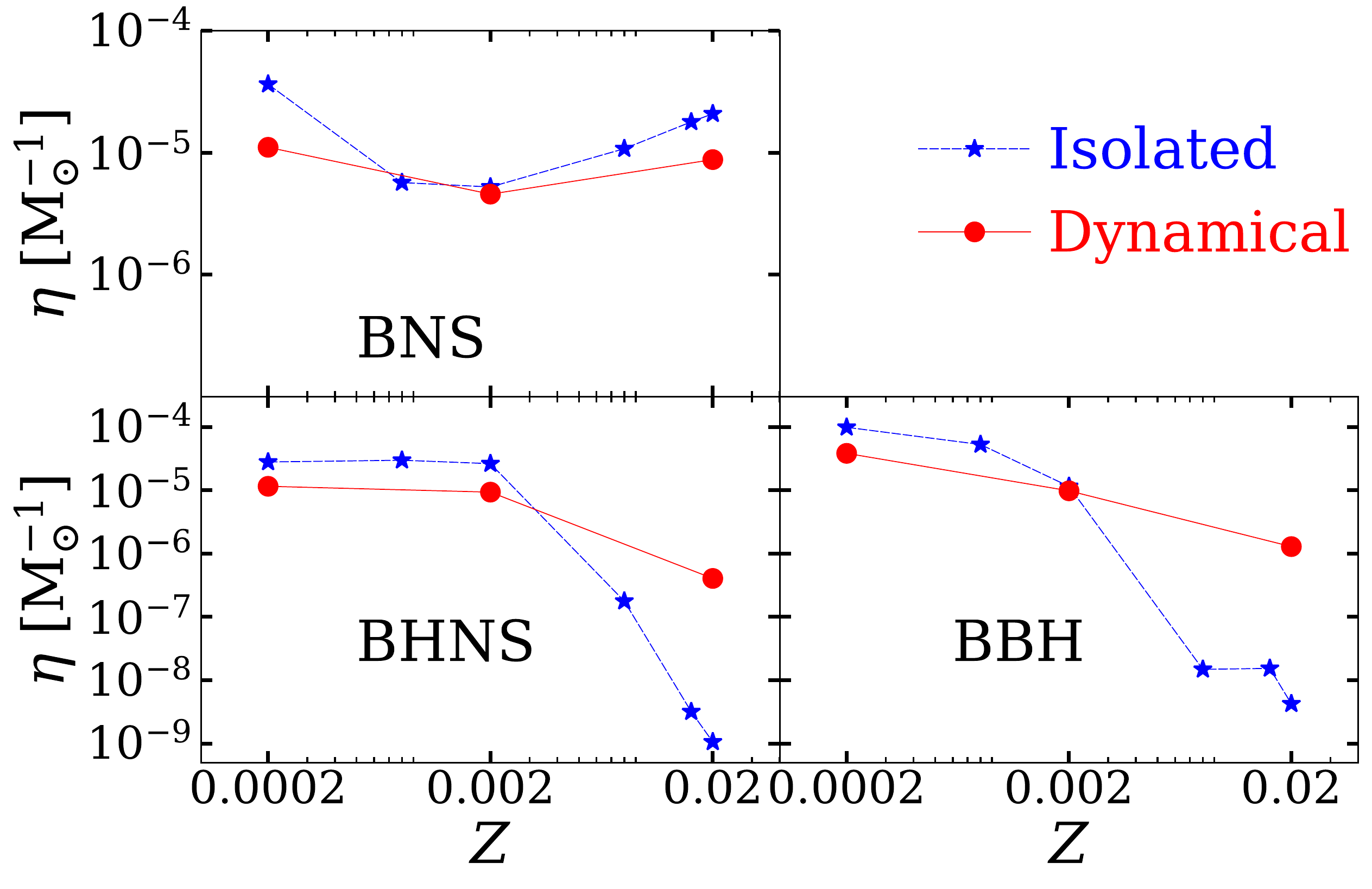}
\caption{Merger efficiency ($\eta$) as a function of progenitor’s metallicity for binaries formed in isolation (blue dashed line and stars) and in young star clusters (red solid line and filled circles).}
\label{fig:eta}
\end{figure}

Only three metallicities ($Z = 0.0002$, 0.002 and 0.02) were available from young star cluster simulations \citep{rastello2020, dicarlo2020}. Running a larger metallicity set is computationally prohibitive. Thus, we linearly interpolated the merger efficiency $\eta{}(Z)$ (Figure~\ref{fig:eta}) in our dynamical simulations to infer the values of $\eta{}(Z)$ for three additional metallicities ($Z = 0.0008$, 0.008, 0.016). 
We assigned to these three interpolated metallicities the available catalogs of dynamical CB mergers with the closest metallicity to the interpolated values. 

\section{Results}

\subsection{Merger efficiency}

 Figure~\ref{fig:eta} shows the merger efficiency $\eta{}(Z)$ from young star clusters and isolated binaries.  This quantity  gives us an idea of the impact of progenitor's metallicity on the merger rate in the different scenarios (isolated and dynamical) we considered. The trend of BNS merger efficiency with metallicity is similar in young star clusters and in isolated binaries, but isolated binaries are more efficient in producing BNS mergers.  The main reason is that dynamical encounters may perturb the evolution of relatively low mass binaries (such as BNSs and their progenitors), widening their orbit or even leading to their disruption (e.g. \citealt{hills1980,ye2020}).
 
 As already noted in several other works (e.g. \citealt{dominik2013,giacobbo2018b,klencki2018,mapelli2019}), the merger efficiency of BNSs is not significantly affected by progenitor's metallicity.

The most interesting difference between isolated binaries and young star clusters is the behavior of BHNSs and BBHs at solar metallicity. The merger efficiency at solar metallicity is about a factor of 100 higher for BBHs/BHNSs formed in young star clusters than for BBHs/BHNSs formed in isolated binaries. The vast majority of dynamical BBH/BHNS mergers at solar metallicity originate from dynamical exchanges\footnote{ Exchanges favor the formation of the most massive binaries in a star cluster \citep{hills1980}. BHs are particularly efficient in acquiring companions through dynamical exchanges, because they are among the most massive objects in a star cluster.} (see \citealt{dicarlo2020} for further details).
 This means that dynamical encounters tend to boost the merger rate of BBHs and BHNSs in the solar metallicity environment.

\subsection{Cosmic MRD}
\begin{figure}
\includegraphics[width = 0.45 \textwidth]{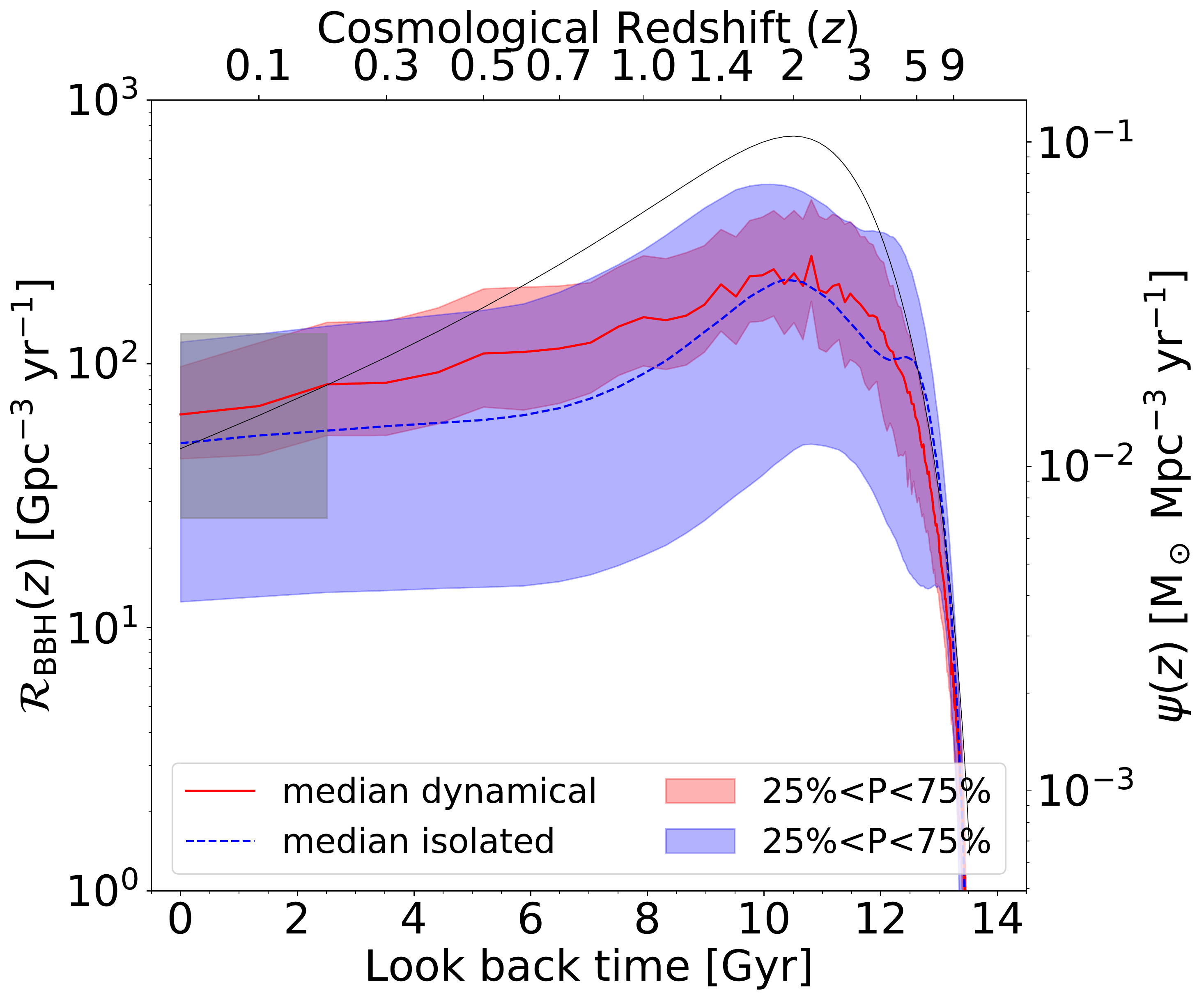}
\caption{The thick lines show the evolution of the MRD of BBHs $\mathcal{R}_{\rm BBH}(z)$ in the comoving frame, calculated as explained in section \ref{sec:mrd}, for BBHs that form in young star clusters (red solid line) and isolated binaries (blue dashed line). The shaded areas represent 50\% of all realizations (between the 75\% percentile and the 25\% percentile). The black solid thin line is the SFR density (from equation \ref{eq:sfrd}). The gray shaded area shows the 90\% credible interval for the local BBH MRD, as inferred from the LVC  \citep{abbottO2,abbottO2popandrate}. The width of the gray shaded area on the $x-$axis  corresponds to the instrumental horizon obtained by assuming BBHs of mass $(10+10)$ M$_\odot$ and O2 sensitivity \citep{abbott2018observingscenario}. 
} 
\label{fig:mrd_bbhs}
\end{figure}

Figure~\ref{fig:mrd_bbhs} shows the MRD of BBHs as a function of time when considering young star clusters (i.e. dynamical binaries) and isolated binaries. In either case, we assume that the entire population of mergers forms from a single channel (i.e. either from young star clusters or from isolated binaries). It is more likely that a percentage of all mergers comes from young star clusters and another percentage from isolated binaries. In a follow-up paper (Bouffanais et al., in prep), we will try to constrain these percentages based on current LVC results. Here, we just want to compare the differences between the two scenarios.

The MRD of BBHs (in both young star clusters and isolated binaries) grows with redshift (a MRD uniform in comoving volume would be an horizontal line in the plot), peaks at $z\sim{}1.5-2.5$, and finally drops  at $z>2.5$. This trend is mostly determined by the cosmic SFR density, which peaks at $z\sim{}2$, convolved with the delay time and the metallicity dependence. These results are fairly consistent with previous papers, which consider different population-synthesis models, metallicity evolution and SFR evolution with redshift (e.g. \citealt{dominik2013,belczynski2016,mapelli2017,mapelli2018,artale2019a,neijssel2019,tang2020}).

At $z=0$, the median values of the MRD of BBHs formed dynamically in young star clusters (hereafter, dynamical BBHs) and the one of isolated BBHs are $R_{\rm BBH}\sim{}64$ and $50$ Gpc$^{-3}$ yr$^{-1}$, respectively. Both values are consistent with the ones inferred from O1 and O2 \citep{abbottO2popandrate}. The median merger rate of dynamical BBHs is higher than the one of isolated BBHs up to $z\sim{}4$ (see Table~\ref{tab:mrds} for more details). This trend can be interpreted by looking at the merger efficiency (Figure~\ref{fig:eta}): around solar metallicity, the dynamical channel is more efficient than the isolated channel. Hence, we expect a higher number of dynamical BBH mergers with short delay time in the local Universe, where metallicity is higher. In contrast, the merger efficiency of dynamical BBHs formed from metal-poor stars ($Z=0.002$) is a factor of $\sim{}2$ lower than the one of isolated BBHs with the same metallicity. Hence, isolated binaries are associated with a higher merger rate from very metal-poor systems.

The MRD of isolated BBHs increases by a factor of $\sim{}{1.8}$ from local Universe up to $z = 1$, and then it grows up faster from redshift $z = 1$ to redshift $z = 2$ (Table~\ref{tab:mrds}). 
On the other hand, the MRD of dynamical BBHs increases almost with the same trend from $z=0$ to $z\sim{}2$ (i.e. without a change of slope at redshift $z\sim{}1$). The main reason for the change of slope in the MRD of isolated BBHs is again the stronger dependence of the merger efficiency on metallicity. In the isolated model, most mergers at redshift $z<1$ are due to BBHs that formed at higher redshift in lower metallicity environments ($Z\sim{}0.0002$) and have a long delay time \citep{mapelli2017,mapelli2018b}.

The uncertainty on MRD resulting from cosmic SFR and metallicity evolution is large, especially for the isolated scenario. For isolated BBHs, the $50\%$ credible interval spreads over more than one order of magnitude between redshift 0 and 4. The 50\% credible interval for the MRD of dynamical BBHs is contained within the credible interval of isolated BBHs. The 50\% credible interval is smaller for dynamical BBHs, because the merger efficiency is less sensitive to metallicity in the dynamical scenario than in the isolated one (Figure~\ref{fig:eta}).

\begin{figure}
\includegraphics[width = 0.45 \textwidth]{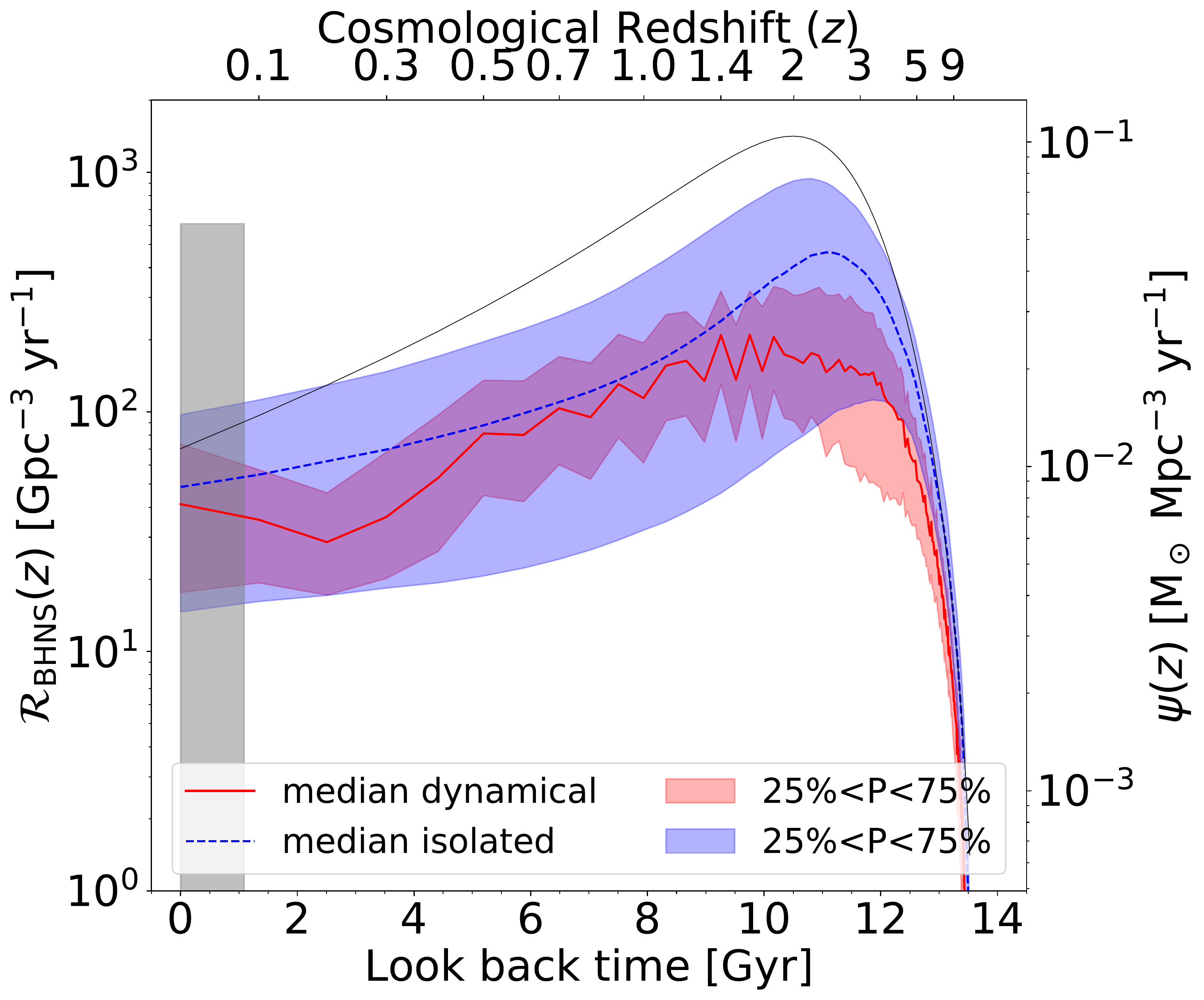}
\caption{Same as Figure~\ref{fig:mrd_bbhs} for BHNSs. The gray box is the upper limit inferred from LVC data \citep{abbottO2}. The width of the gray shaded area on the $x-$axis  corresponds to the instrumental horizon obtained by assuming BHNSs of mass $(1.4+5)$ M$_\odot$ and O2 sensitivity \citep{abbott2018observingscenario}.}
\label{fig:mrd_bhns}
\end{figure}

\begin{figure}
\includegraphics[width = 0.45\textwidth]{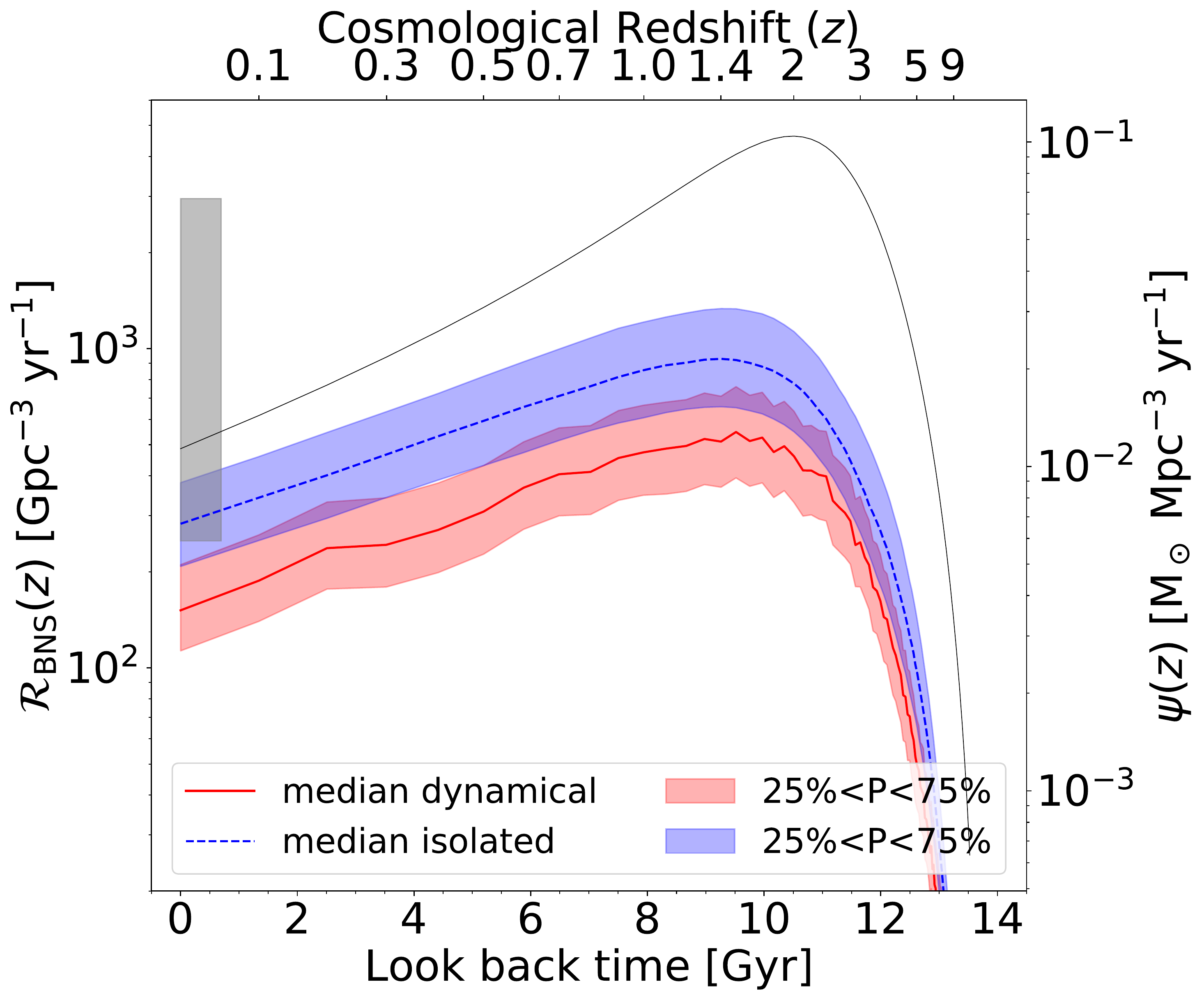}
\caption{Same as Figure~\ref{fig:mrd_bbhs} for BNSs. The gray box is the 90\% credible interval inferred by considering both GW170817 and GW190425 \citep{abbottGW190425}.  The width of the gray shaded area on the $x-$axis  corresponds to the instrumental horizon obtained by assuming BNSs of mass $(1.4+1.4)$ M$_\odot$ and O2 sensitivity \citep{abbott2018observingscenario}.}
\label{fig:mrd_bnss}
\end{figure}

Figure~\ref{fig:mrd_bhns} shows the MRD evolution of BHNSs. At $z=0$, $\mathcal{R}_{\rm BHNS}= 41^{+33}_{-23}$ and $49^{+48}_{-34}$ Gpc$^{-3}$ yr$^{-1}$  for dynamical and isolated BHNSs, respectively. At redshift $z=2$, $\mathcal{R}_{\rm BHNS}=168^{+138}_{-76}$ and $406^{+516}_{-331}$ Gpc$^{-3}$ yr$^{-1}$  for dynamical and isolated BHNSs, respectively.  
For most of the cosmic time, the boundaries of the $50\%$ credible intervals of our two models have similar values. 
The higher boundary of the 50\% credible interval  for both dynamical and isolated BHNSs is below the upper limit from the LVC ($\mathcal{R}_{\rm BHNS}<610$ Gpc$^{-3}$ yr$^{-1}$, \citealt{abbottO2}), indicating that our model is consistent with O1 and O2 results. 
In the case of both BBHs and BHNSs, most of the uncertainty comes from metallicity evolution, because BBHs and BHNSs are extremely sensitive to metallicity variations (as shown in Figure~\ref{fig:eta}).

 Finally, Figure~\ref{fig:mrd_bnss} shows the MRD evolution of dynamical and isolated BNSs. At redshift $z\leq{}0.1$, the MRD of dynamical BNSs ($\mathcal{R}_{\rm BNS}=151^{+59}_{-38}$ Gpc$^{-3}$ yr$^{-1}$) is a factor of $\sim{}2$ lower than the one of isolated BNSs ($283^{+97}_{-75}$ Gpc$^{-3}$ yr$^{-1}$). A similar difference is found at $z=2$, where the MRD is  $\mathcal{R}_{\rm BNS}=460^{+177}_{-130}$ and $777^{+354}_{-228}$ Gpc$^{-3}$ yr$^{-1}$, for dynamical and isolated BNSs respectively. Overall, the MRD of dynamical BNSs is significantly lower than the one of isolated BNSs, even if the MRD evolution with redshift is similar. 
This trend is expected by looking at Figure~\ref{fig:eta},   
because the merger efficiency  of dynamical BNSs is lower at all metallicities. In young star clusters, the formation of BNSs is slightly suppressed with respect to isolated binaries, because such relatively low-mass binaries tend to be broken or softened (i.e. their orbital separation is increased) by dynamical encounters.

The local MRD of isolated BNSs is consistent with the one inferred from the LVC, while the local MRD of dynamical BNSs is below the 90\% credible interval from the LVC.  This suggests that (young) star clusters {\emph{alone}} might not be able to explain all the BNS mergers detected by the LVC.

The models presented in this work assume small natal kicks for neutron stars, which are in tension with the proper motions of Galactic young pulsars \citep{giacobbo2018b}. We recently proposed a new model for natal kicks that can reproduce the proper motions of Galactic pulsars and  gives a value for the MRD close to the one presented in this study \citep{giacobbo2020}. As a result, we do not expect significant differences in the MRD between the model adopted in this work and the one proposed by \cite{giacobbo2020}. 
 
The 50\% credible interval of simulated BNSs is significantly smaller than that of both BHNSs and BBHs, because BNSs are less sensitive to stellar metallicity (Fig.~\ref{fig:eta}). Hence, the uncertainty on BNS merger rate comes mostly from the SFR, for a fixed binary evolution model.
 
 Our local MRDs for dynamical BNSs and BHNSs are higher than the values estimated by \cite{ye2020} for globular clusters ($\mathcal{R}_{\rm BNS}\sim{}\mathcal{R}_{\rm BHNS}\sim{}0.02$ Gpc$^{-3}$ yr$^{-1}$). This is not surprising because globular clusters form mostly at $z\gtrsim{}2$, while smaller star clusters, like the ones we simulated, form all the time from high to low redshift and are an important channel of star formation in the local Universe.
\renewcommand{\arraystretch}{1.5}
\begin{table*}
\begin{center}

    \caption{MRD in $[\text{Gpc}^{-3}\,{}\text{yr}^{-1}]$ for five redshift intervals. We show a comparison between dynamical CBs formed in young star clusters and  isolated CBs.}    
\begin{tabular}{lcccccccccc}
    \hline
    \hline
         &  \multicolumn{10}{c}{Redshift intervals}\\
     
         & \multicolumn{2}{c}{$z \in [0,0.1]$} &   \multicolumn{2}{c}{$z \in [0.9,1.0]$} &  \multicolumn{2}{c}{$z \in [1.9,2.0]$} &  
         \multicolumn{2}{c}{$z \in [2.9,3.0]$} &
         \multicolumn{2}{c}{$z \in [3.9,4.0]$}   \\

        CB & Dynamical & Isolated &  Dynamical & Isolated &  Dynamical & Isolated &   Dynamical & Isolated &  Dynamical & Isolated \\
          \hline

          BBH & $64^{+34}_{-20}$ & $50^{+71}_{-37}$ & $150^{+107}_{-52}$& $92^{+178}_{-73}$ & $220^{+161}_{-77}$ & $207^{+256}_{-160}$ &$168^{+136}_{-71}$ & $130^{+192}_{-91}$ & $101^{+75}_{-51}$ & $105^{+191}_{-83}$\\
          BHNS & $41^{+33}_{-23}$ & $49^{+48}_{-34}$ & $114^{+80}_{-53}$ & $152^{+227}_{-120}$ & $168^{+138}_{-76}$ & $406^{+516}_{-331}$ &$142^{+129}_{-91}$ & $395^{+286}_{-286}$ & $99^{+62}_{-55}$ & $225^{+131}_{-124}$\\
          BNS & $151^{+59}_{-38}$ & $283^{+97}_{-75}$ &$473^{+192}_{-126}$ &$856^{+355}_{-249}$ &$460^{+177}_{-130}$ & $777^{+354}_{-228}$ & $247^{+98}_{-68}$ & $379^{+191}_{-113}$ &$110^{+44}_{-31}$ & $190^{+98}_{-63}$ \\
          \hline
    \end{tabular}
    \label{tab:mrds}
\end{center}
\end{table*}

\subsection{Mass distribution}
\begin{figure}
  \centering
		\subfigure{
		\includegraphics[width=0.45\textwidth]{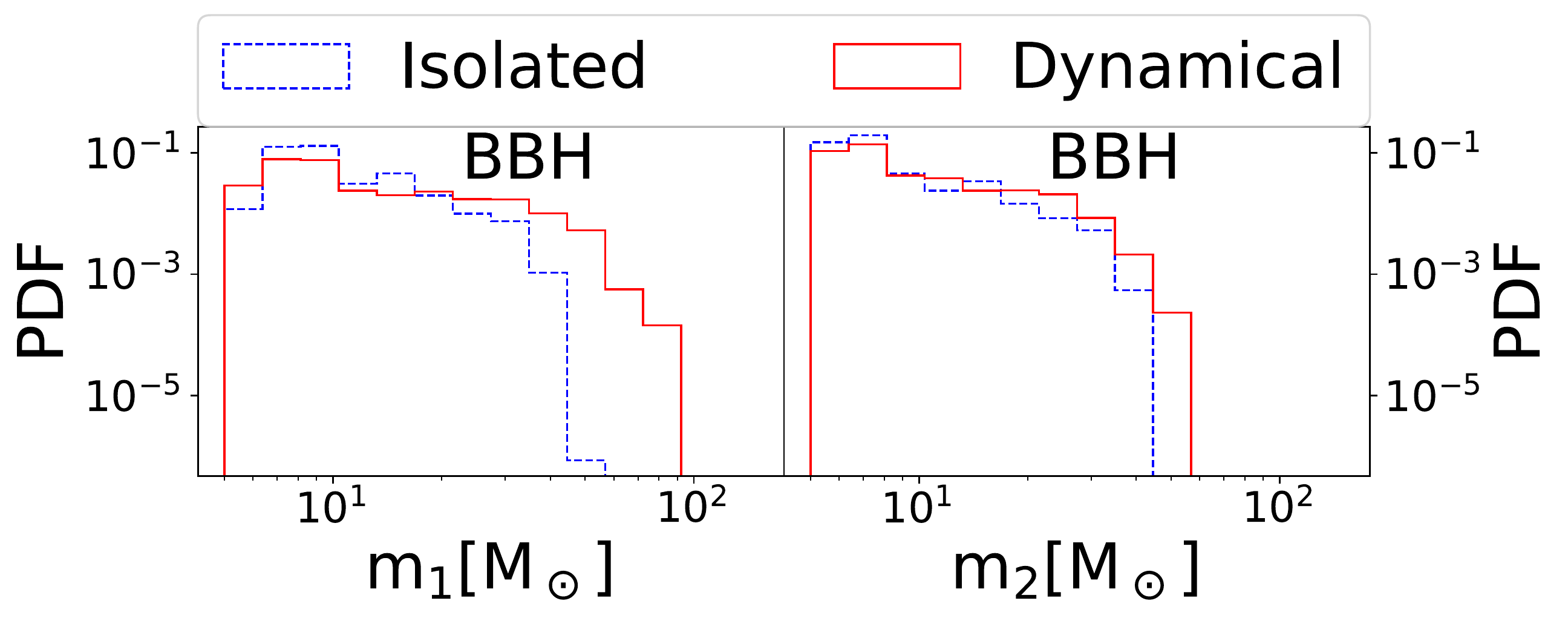}}\vspace*{-2.45em}\\
		\subfigure{
		\includegraphics[width=0.45\textwidth]{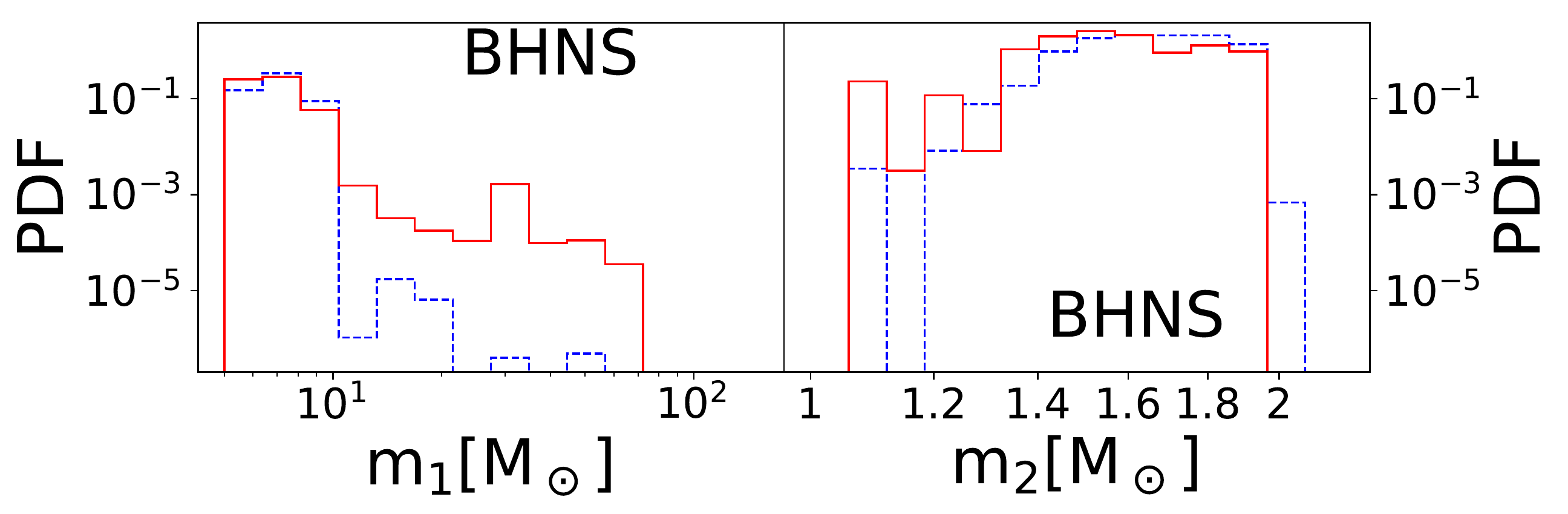}}\vspace*{-2.45em}\\
		\subfigure{
		\includegraphics[width=0.45\textwidth]{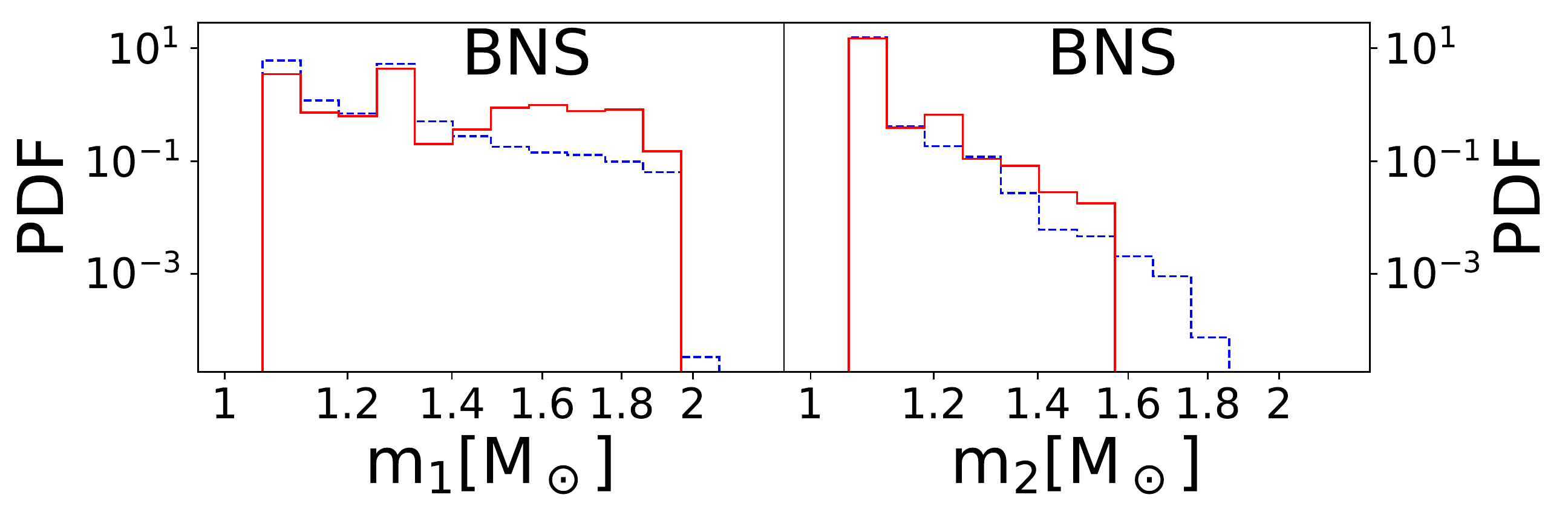}}\\
  \caption{Distribution of primary (left) and secondary mass (right) of BBHs (top), BHNSs (middle) and BNSs (bottom panel).  Blue dashed and red solid histograms refers to isolated and dynamical CBs, respectively}.\label{fig:mass}
\end{figure}

Figure~\ref{fig:mass} shows the mass distribution of BBHs, BHNSs and BNSs merging across cosmic time. We plot together binaries merging at different redshift because we find no significant dependence of the mass distribution on the merger redshift, consistent with \cite{mapelli2019}. The main difference between the mass distribution of dynamical BBHs and the one of isolated BBHs is that low-mass BBHs are less numerous in the former than in the latter scenario. Moreover, the maximum mass of merging BHs from isolated binaries is $m_{\rm BH,\,{}max}\sim{}45$ M$_\odot$, whereas dynamics in young star clusters leads to a significantly larger maximum mass $m_{\rm BH,\,{}max}\sim{}90$ M$_\odot$. Quantitatively, the percentage of isolated BBHs that have a primary mass $> 40$ M$_\odot$ is equal to 0.07\%, while it is 10.6\%  for dynamical BBHs. This marked difference in the maximum mass of merging BHs between isolated and dynamical BBHs can be understood as follows (see also \citealt{dicarlo2019a,dicarlo2019b}). The stellar wind and core collapse SN prescriptions  adopted in {\sc mobse} allow the formation of BHs with mass up to $\sim{}65$ M$_\odot$ \citep{giacobbo2018}, but only BHs with masses up to $\sim{}45$ M$_\odot$ are able to merge within a Hubble time in isolated BBHs, because of a subtle interplay between mass transfer and stellar radii. In fact, BHs with masses $>45$ M$_\odot$ form only from stars with zero-age main sequence mass $\sim{}60-80$ M$_\odot$  which retain a large fraction of hydrogen envelope and collapse to a BH directly (Figure~4 of \citealt{giacobbo2018}). When such stars are members of a tight binary system, most of the hydrogen envelope is removed by mass transfer (or by common envelope) before the collapse; hence, even if they might end up into a BBH merger, the mass of the final BHs will be smaller than the one we expect from single star evolution. In contrast, if such stars are members of loose binaries (initial orbital separation $a\gtrsim{}10^4$ R$_\odot$), which do not undergo mass transfer, they produce BBHs with individual BH masses $>45$ M$_\odot$, but the orbital separation is too large to lead to coalescence.

In young star clusters, instead, BHs with masses $>45$ M$_\odot$ are able to merge within a Hubble time, because i) if they form from the collapse of single stars, they can acquire companions through dynamical exchanges, and ii) if they are members of loose binaries, these massive binaries are efficiently hardened by three body encounters \citep{dicarlo2019a}.  Moreover,  (multiple) stellar mergers can even lead to the formation of BHs with masses $>65$ M$_\odot$, as discussed in \cite{dicarlo2019b}. Such massive BHs are single at birth but can acquire a companion by dynamical exchanges.

Figure~\ref{fig:mass} shows that dynamical BHNSs can host significantly more massive BHs than isolated BHNSs. 
Only $9\times{}10^{-4}\%$  
of BHs in isolated BHNSs have masses $m_{\rm BH}>20$ M$_\odot$, while 1.6\% 
of BHs in dynamical BHNSs have masses above this value.  This is another effect of dynamics, which boosts the formation of massive binaries by dynamical exchanges and facilitates the coalescence of binaries with extreme mass ratio by dynamical hardening (see the discussion in \citealt{rastello2020} for additional details). Finally, we do not find any significant difference between the mass distribution of dynamical BNSs and that of isolated BNSs\footnote{The cut-off of secondary NS masses above $\sim{}1.6$ M$_\odot$ in the dynamical model is a consequence of the lower statistics of dynamical BNSs with respect to isolated BNSs in the original catalogs we used.}.

\section{Summary}
The next generation of ground-based GW interferometers (Einstein Telescope and Cosmic Explorer) will observe BBH (BNS) mergers up to $z\gtrsim{}10$ ($z\sim{}2$), allowing us to probe the evolution of CBs across cosmic time. Here, we have investigated the cosmic evolution of CBs formed in young star clusters by evaluating their MRD. Young star clusters are the most common birthplace of massive stars across cosmic history. 
Hence, a large fraction of BBHs, BHNSs and BNSs might have formed in young star clusters and might retain the signature of dynamical processes (such as exchanges or stellar collisions) occurring in star clusters.

The dynamical BBH merger rate is higher than the isolated BBH merger rate between $z=0$ and $z\sim{}4$. The main reason for this difference is that the merger efficiency of dynamical BBHs at solar metallicity is two orders of magnitude higher than the merger efficiency of isolated BBHs, because dynamical exchanges enhance the merger of BBHs formed from metal-rich stars.

 The MRD of dynamical BHNSs  is always consistent with that of isolated BHNSs, within the estimated uncertainty. In contrast, the MRD of dynamical BNSs is a factor of $\sim{}2$ lower than that of isolated BNSs, because dynamics suppresses the formation of relatively low-mass binaries. 

We find a local MRD of $\mathcal{R}_{\rm BBH}=64^{+34}_{-20}~\text{Gpc}^{-3}\text{yr}^{-1}$, $\mathcal{R}_{\rm BHNS}=41^{+33}_{-23}~\text{Gpc}^{-3}\text{yr}^{-1}$ and $\mathcal{R}_{\rm BNS}=151^{+59}_{-38}~\text{Gpc}^{-3}\text{yr}^{-1}$ for dynamical BBHs, BHNSs and BNSs, respectively. The rates of dynamical BBHs and BHNSs are consistent with the values inferred from O1 and O2 \citep{abbottO2,abbottO2popandrate} within the uncertainties, while the rate of dynamical BNSs is below the lower edge of the 90\% credible interval inferred  by the LVC ($250-2810$ Gpc$^{-3}$ yr$^{-1}$, \citealt{abbottGW190425}).  The local MRDs of isolated BBHs, BHNSs and BNSs ($\mathcal{R}_{\rm BBH}=50^{+71}_{-37}$ Gpc$^{-3}$ yr$^{-1}$, $\mathcal{R}_{\rm BHNS}=49^{+48}_{-34}$ Gpc$^{-3}$ yr$^{-1}$ and $\mathcal{R}_{\rm BNS}=283^{+97}_{-75}$ Gpc$^{-3}$ yr$^{-1}$) are all consistent with the values inferred from O1 and O2.

The main difference between isolated BBHs/BHNSs and dynamical BBHs/BHNSs is the mass of the BH component: dynamical systems harbor BHs with mass up to $m_{\rm BH,\max{}}\sim{}90$ M$_\odot$, significantly higher than isolated binaries ($m_{\rm BH,\,{}max}\sim{}45$ M$_\odot$). The mass distribution of both isolated and dynamical CBs do not significantly change with redshift. These results provide a clue to differentiate the dynamical and isolated formation scenario of binary compact objects  across cosmic time, in preparation for next-generation ground-based detectors.

\section*{acknowledgments}
We thank the anonymous referee for their useful comments. We are also grateful to Marica Branchesi, Guglielmo Costa, Mario Pasquato, Giuliano Iorio and Stefano Torniamenti for useful discussions.  MM, FS, NG, YB, SR and AB acknowledges financial support by the European Research Council for the ERC Consolidator grant DEMOBLACK, under contract no. 770017. MCA and MM acknowledge financial support from the Austrian National Science Foundation through FWF stand-alone grant P31154-N27 ``Unraveling merging neutron stars and black hole -- neutron star binaries with population synthesis simulations''.

\vspace{5mm}
\software{ {\sc mobse} \citep{giacobbo2018}; {\sc nbody6++gpu} \citep{wang2015};
{\sc cosmo$\mathcal{R}$ate} (this paper).}






\bibliography{santoliquido}{}
\bibliographystyle{aasjournal}



\end{document}